\documentclass[journal]{IEEEtran}
 \usepackage{mathtools} 
 \usepackage{amsfonts}
 \usepackage{amsmath}
 \usepackage{amsthm}
 \usepackage{amssymb}

\usepackage{algorithm}
\usepackage{algorithmicx}
\usepackage{algpseudocode}
\usepackage{booktabs}
\usepackage{enumitem}
\usepackage{subfigure}
\usepackage{graphicx}
\usepackage{balance}
\usepackage{xcolor}
\usepackage{url}
\usepackage{cite}  
\usepackage{soul}

\ifCLASSINFOpdf
\else
\fi

\begin{document}
    \title{6G Wireless Communications in 7--24 GHz Band: Opportunities, Techniques, and Challenges}
\author{Zhuangzhuang~Cui,~Peize~Zhang,
and Sofie Pollin
\thanks{Zhuangzhuang Cui and Sofie Pollin are with the WaveCoRE of the Department of Electrical Engineering (ESAT), KU Leuven, Belgium (e-mail:\{zhuangzhuang.cui,~sofie.pollin\}@kuleuven.be).}
\thanks{Peize Zhang~is with the Centre for Wireless Communications, University of Oulu, Finland (e-mail: peize.zhang@oulu.fi).}
}

\maketitle

\begin{abstract}
The sixth generation (6G) wireless communication nowadays is seeking a new spectrum to inherit the pros and discard the cons of sub-6 GHz, millimeter-wave (mmWave), and sub-terahertz (THz) bands. To this end, an upper mid-band with a Frequency Range (FR) spanning from 7 GHz to 24 GHz, also known as FR3, has emerged as a focal point in 6G communications. Thus, as an inevitable prerequisite, a comprehensive investigation encompassing spectrum utilization and channel characteristics is the first step to exploiting potential applications and prospects of using FR3 in the 6G ecosystem. In this article, we provide FR3 synergies with emerging technologies including non-terrestrial network (NTN), massive multi-input multi-output (mMIMO), reconfigurable intelligent surface (RIS), and integrated sensing and communications (ISAC). Furthermore, leveraging ray-tracing simulations, our investigation unveils the similarity of channel characteristics in FR3 with other FRs. The analysis of RIS-aided communication shows the insight of higher spectral efficiency achieved in FR3 compared to other FRs when using the same RIS size. Finally, challenges and promising directions are discussed for wireless systems in FR3.  
\end{abstract}

\IEEEpeerreviewmaketitle

\section{Introduction}
In the sixth generation (6G) mobile communication, different parts of the radio spectrum are used, like sub-6 GHz and millimeter-wave bands. These bands are specialized for 5G New Radio (NR) as Frequency Range (FR) 1 and 2~\cite{3gppfr1}. In recent years, sub-terahertz (THz) has become increasingly popular for 6G in providing high-accurate sensing and extreme-fast transmission. 
Nonetheless, systems using higher frequencies tend to cover less area because of how radio waves propagate. Therefore, the most valuable new spectrum is expected to balance coverage, capacity, and deployment challenges for typical scenarios, and offer strong benefits when considering new trends such as non-terrestrial network (NTN), reconfigurable intelligent surface (RIS), and integrated sensing and communications (ISAC). As a result, the upper mid-band has attracted large interest from both academia and industry, as evidenced by Nokia \cite{nokiabell}. Recently, World Radiocommunication Conference 2023 (WRC-23) has studied new frequency bands above 6~GHz for international mobile telecommunications (IMT)~\cite{wrc23final}, in which 6.425--7.125~GHz and 10--10.5 GHz (with condition) were identified, and the other two ranges including 7.125--8.4~GHz and 14.8--15.35~GHz were potentially identified and will be studied at WRC-27.  
These regulations open the door for the 6G wireless systems operating in the 7--24 GHz band, resulting in consecutive microwave connectivity by efficiently manipulating among sub-6~GHz, 7--24 GHz, mmWave, and sub-THz.

The 3rd Generation Partnership Project (3GPP) Release 16 initiated the investigation of NR in 7--24 GHz~\cite{3gppfr3}, in which a comprehensive regulatory landscape survey on this FR is provided based on WRC and International Telecommunication Union (ITU). It shows that the spectrum potentially allocated 
to mobile service on a primary basis includes 7.125--8.4, 10--10.5, 10.7--13.25, 14.8--15.35, 17.7--19.7, and 21.2--23.6~GHz, accumulating in an aggregate bandwidth of approximately 9 GHz. Currently, the ongoing Release 19 has initiated a study item description (SID), targeting channel modeling in FR3. 


Since FR3 is intermediate between FR1 and FR2, an open issue lies in how similar channel behaviors in this FR are compared to the other two, and how the pros and cons of both bands are balanced. 
Overall, the availability of channel measurements 
in this FR is still limited. 
In \cite{kang2023cellular}, the authors used ray-tracing channel simulations to show the favorable coverage performance of cellular networks using FR3. Moreover, they investigated penetration losses in sub-6 GHz, 7-24 GHz, and mmWave up to 60 GHz, which shows the benefit of FR3 in outdoor-to-indoor scenarios. 
In \cite{edgar16}, satellite-to-ground channel measurements were conducted at 11 GHz and 14 GHz, where frequency dispersion is characterized. Despite all that, a comparative study of different FRs is required. 

Regarding the applications in FR3,
ITU outlines major applications in FR3, which reside in earth-space connectivity, deep space, unmanned aerial systems (UAS), and radiolocation. It suggests that NTNs have become promising applications, where commercial endeavors like Starlink, functioning within 10.7--12.7 GHz, have demonstrated the viability of FR3. For terrestrial networks, a study from Nokia shows that using 3.5 GHz and 28 GHz can provide an urban cell range of 430~m and 190~m, respectively, while employing 7–14~GHz provides 260–375~m coverage ranging from rooftop to indoor \cite{nokiabell}. FR3 indeed provides 
a realistic coverage compromise. However, the cell range of FR3 shows better urban macro connectivity with continuous outdoor-to-indoor (O2I) coverage, which is difficult to achieve in mmWave bands.  
\begin{figure*}[!t]
\centering
    \includegraphics[width=6in]{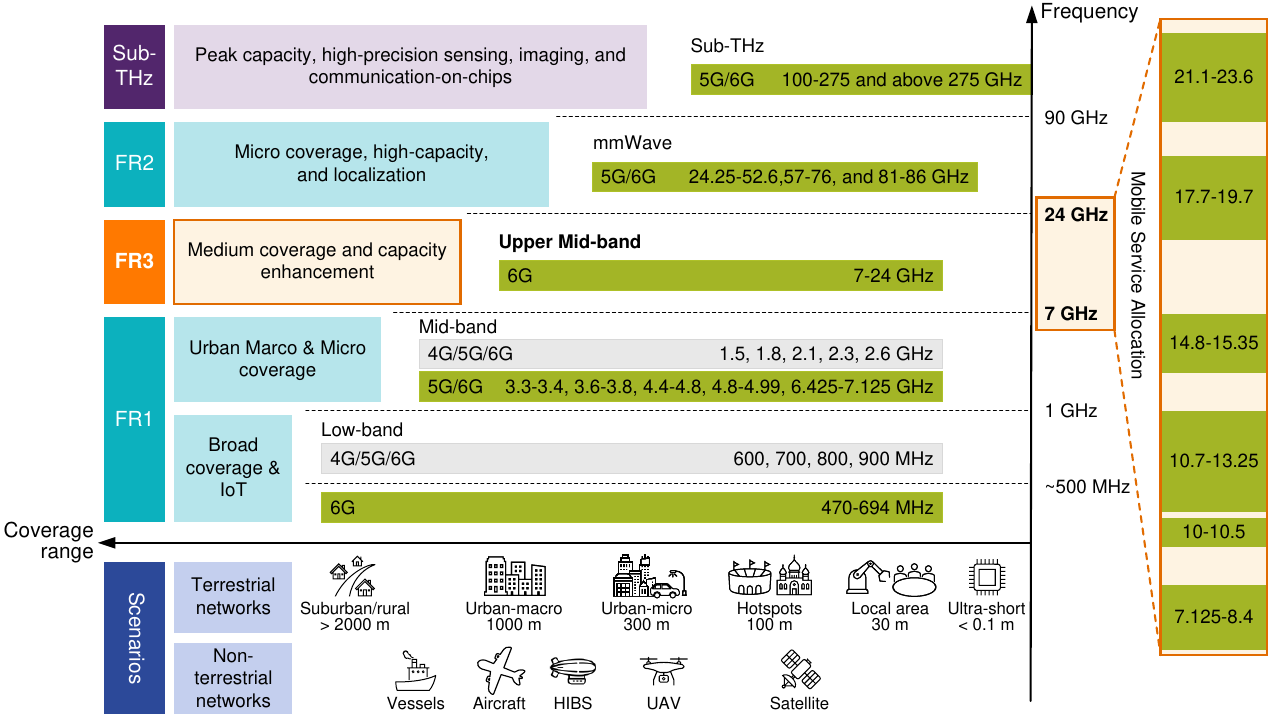}
    \caption{Spectrum overview for mobile networks
    .}
    \label{Fig:Spectrum}
\vspace{-6pt}
\end{figure*}

When considering multiple-input multiple-output (MIMO) systems, FR3 brings unique advantages in terms of number of antennas versus multipath opportunities. As a result, trends such as extremely large Massive MIMO and even large Reconfigurable Intelligent Surfaces (RIS) become very realistic at that frequency \cite{antennafr3}, giving rise to unique new research opportunities. Indeed, an FR3 system has the opportunity to migrate from a rich multi-path small cell with extreme spatial multiplexing (serving a hot spot) to an extremely large array with sharp beamforming towards aerial nodes or even deep space. 
One fact is the considerably smaller wavelength in this FR (e.g., 15-24 GHz has a wavelength of 1.99-1.25 cm) compared to the minimum 5~cm at sub-6 GHz, which helps add a larger number of antenna elements within the same area, facilitating space-limited nodes such as satellite and UAVs. Moreover, the hardware design in this FR has been studied for decades such as 0.13-$\mu$m CMOS 6-bit active digital phase shifters at $K_u$-band \cite{duan18}. These existing works evidence the feasibility of the practical use of FR3 in multi-antenna systems. 
Another facet pertains to the widespread implementation of radar technology throughout the entirety of the FR3. In \cite{3gppfr3}, it is found that synthetic aperture radar and weather radar employ $X$ band and maritime radar can be deployed at $K_u$ band. Moreover, radar at $K$ bands is widely used for detecting drones. These off-the-shelf deployments provide legacies for ISAC applications in FR3. 

This article will detail the opportunities of FR3. 
The properties of radio propagation in different FRs will then be studied using ray tracing simulations and a comprehensive overview of relevant channel parameters. 
Following that, we will elaborate on the benefits when employing FR3 in emerging techniques. Finally, we will provide the challenges of FR3 implementation.

\begin{table*}[t]
\caption{A comparison among different FRs.}
\begin{center}
\begin{tabular}{|c|c|c|c|c|c|c|c|}
\hline
\textbf{FR}& \textbf{Range} & \textbf{Bandwidth}~\cite{3gppfr3}  & \textbf{Suggested scenario} & \textbf{Coverage} & \textbf{Throughput} & \textbf{Hardware cost} & \textbf{Array directivity$^{\mathrm{a}}$ (HPBW)} \\
\hline
FR1& 450 MHz -- 6 GHz & 5--100 MHz&  UMa, RMa, UMi & Wide & Low & Low & 11.72~dBi $@$3.5 GHz ($50.6^\circ$)\\
\hline
FR2& 24.25--52.6 GHz & 50--400 MHz& UMi, InH, Backhaul & Narrow & High & High & 27.79~dBi $@$28 GHz  ($7.9^\circ$)\\
\hline
FR3& 7.125--24.25 GHz& 5--400 MHz&  RMa, UMi, O2I, InH & Medium & Flexible & Medium &  21.60~dBi $@$14 GHz ($16.2^\circ$) \\
\hline
\multicolumn{7}{l}{$^{\mathrm{a}}$ We consider a half-wavelength-spaced URA with the same array aperture of $8.56 \times 8.56$~cm$^2$.}
\end{tabular}
\label{tab1}
\end{center}
\end{table*}

 \section{FR3 Opportunities 
}
This section begins with an overview of the potential spectrum for future 6G. Then, the advantages of exploring FR3 and typical 6G use scenarios are discussed.

\subsection{Radio Spectrum Overview}

In Fig.~\ref{Fig:Spectrum}, we unfold the spectrum utilization in four domains including \textit{frequency, scenario, application, and FR in standardization}. We summarize the main frequency bands used in 4G-5G, and 
expect 
that 6G will be able to employ all allocated frequency bands from 470~MHz to sub-THz synergistically and opportunistically. Scenarios categorized by coverage range, span from broad-covered rural environments to ultra-short on-chip boards. Then, 
application diversity is based on frequency and 
bandwidth options, where low-frequency bands with limited bandwidth generally enable broad coverage such as Internet of Things applications, and high-frequency bands with huge bandwidth will support, e.g., very high throughput and high-precision sensing. Thus, 
considering a multi-service, multi-band, and multi-scenario 6G network, it requires spectrum diversity and corresponding 
spectrum scheduling, e.g., sub-6 GHz for coverage, mmWave for sensing, and FR3 for balancing coverage and throughput. Generally, we compare FR3 with the other FRs in Table~\ref{tab1}, where FR3 can provide more flexible throughput, but indeed its coverage is naturally medium due to frequency and channel quality. However, it provides a chance for ISAC when considering the same-size array using different FRs, because the array's half-power beamwidth (HPBW) is narrower than FR1, and propagation conditions are better than FR2. Note that the phased array directivity and HPBW are obtained from Matlab$^\circledR$ simulation. 


\subsection{Advantages of FR3 Band}

FR3 offers distinct advantages compared to FR1, regarding a wider channel bandwidth. Compared with FR2, less signal attenuation and cheaper hardware implementation costs in FR3 are valuable. We highlight its advantages from coverage, flexibility, and functionality perspectives.

\subsubsection{Enhancing 3D Coverage} 
As we typically find higher bandwidths only at higher frequencies, communication systems are designed to deliver coverage only in lower frequencies, and capacity only in higher frequencies, leaving the upper mid-band as unallocated. 
Moreover, distinct channel conditions in sub-6 GHz and mmWave bands lead to 
insufficient flexibility in real deployments. For example, an indoor user generally demands a high throughput, which is hardly met by a sub-6 GHz base station (BS) with a limited bandwidth. MmWave bands become inefficient for outdoor deployment because large signal attenuation requires massive BSs. As there are indoor and outdoor communication needs, this results in a double deployment cost, as both indoor and outdoor coverage and networks will have to be provided in both low and high-frequency ranges. 
FR3, however, strikes a balance in channel fading and bandwidth making it a golden band for \textit{continuous} outdoor-indoor coverage. Moreover, the intermediate wavelength of FR3 enables the usage of massive antennas so that 3D beamforming can be efficiently employed to support \textit{vertical} users such as UAVs. With particular interest, the band of 7.125-8.4 GHz has been identified for study use by the Federal Communications Commission (FCC).  


\subsubsection{Enabling Flexibility} 
6G networks are expanding MIMO 
towards array sizes above 256 antennas in 5G NR. Such a massive MIMO deployment requires a balance between its cost and efficiency. Due to the larger wavelength in FR1, the physical sizes of the extreme-large antenna arrays (ELAAs) will dramatically increase, leading to spatial non-stationarity. Meanwhile, deploying large physical arrays in FR1 is challenging due to greater wind loading and weight 
constraints. In FR2, due to the sparse scattering, favorable propagation effects such as channel hardening no longer exist. Moreover, the implementation complexity and cost for digital beamforming in mmWave bands also hinder the ELAA deployment \cite{mmimofr3}.

In general, FR3 offers the following key advantages for boosting capacity flexibly. Firstly, its centimeter wavelength permits a larger number of antenna elements compared to sub-6 GHz. Secondly, when equipping with a more extensive array, the channel hardening effect becomes more pronounced, which makes the small-scale fading less relevant, i.e., flattening out the channel fading in frequency and time after spatial processing. Channel hardening also enables spatial multiplexing with low-cost linear MIMO processing. 
The other is that FR3 has more favorable channel conditions compared to mmWave, which improves channel rank and condition number, thus better exploiting spatial multiplexing of MIMO systems.
FR3 systems are overall smaller, less complex, and experience better channel hardening, which results in more flexible performance benefits with lower costs.




\subsubsection{Empowering Sensing} 
A prevailing trend in 6G networks involves multi-service and multi-band applications that extend beyond traditional communication-only functionality. One such promising application is ISAC.  
Numerous existing radar studies in $X$, $K_u$, and $K$ bands focus on human/vehicle sensing, where the legacies in terms of hardware and algorithmic implementations can be reusable for ISAC. 
The combination of a large number of antennas, large bandwidth, and decent coverage contribute to an efficient implementation of ISAC with a good angle, range, and detection performance. Multi-band sensing with FR1 and FR2 bands is challenging as the very different frequency and bandwidth ranges result in very severe frequency-dependent channels and Doppler effects. An intermediate FR3 could simplify multi-band sensing, as extreme frequency-dependent Doppler effects and very different propagation conditions are avoided.

\begin{figure}[!t]
\centering
    \includegraphics[width=3.5in]{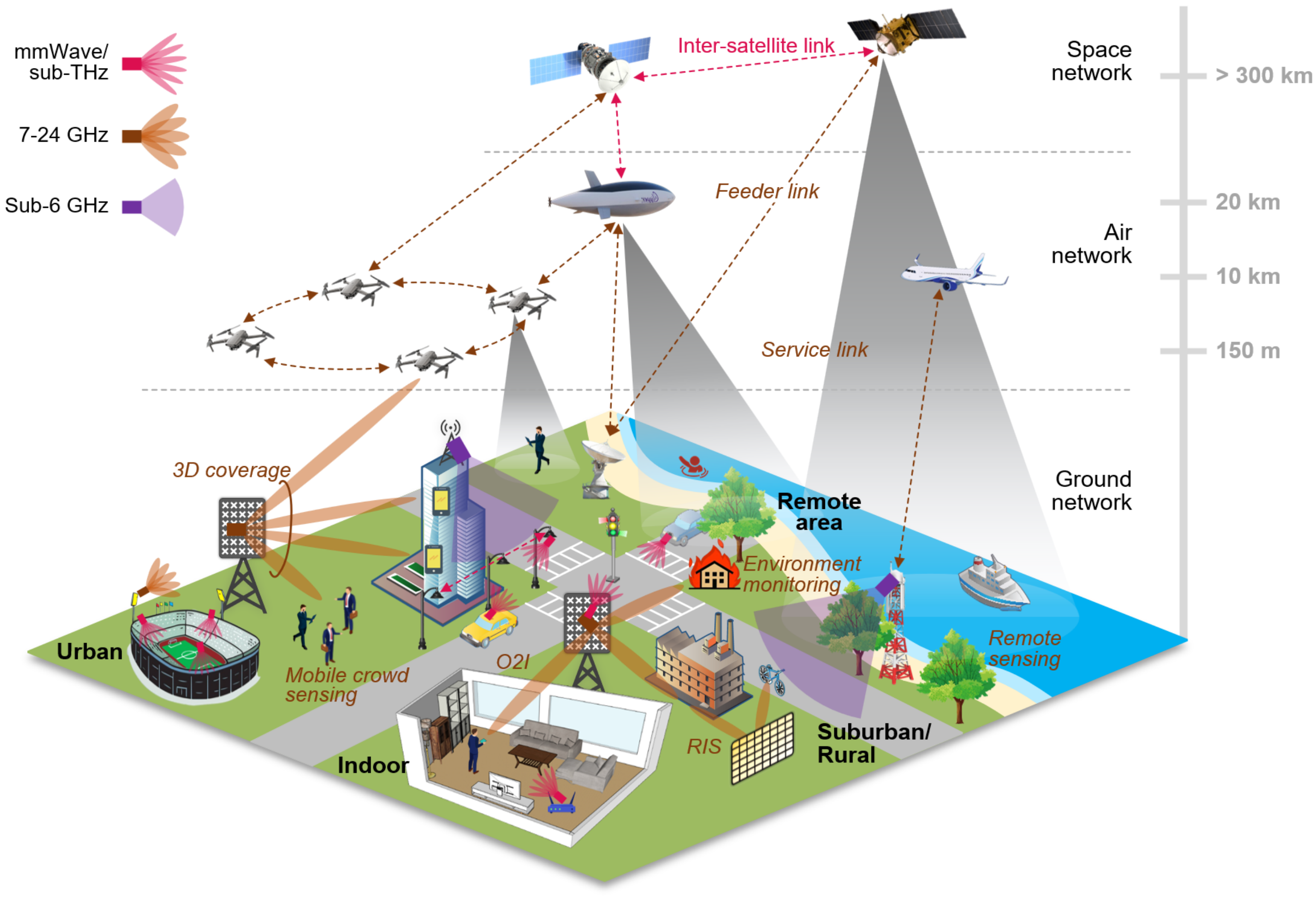}
    \caption{Use cases of FR3 and its coexistence with other FRs in 6G networks.}
    \label{Fig:Usecase}
\vspace{-6pt}
\end{figure}

\subsection{Use Cases}
Given the regulations, existing technologies, and future trends in FR3, the promising use cases include space-air-ground networks with continuous and vertical coverage, multi-antenna systems, and multi-band multi-service applications. Specifically, inter-satellite and satellite-ground links have been deployed at 22.2-23.5 GHz and 10.7--12.7 GHz, respectively. In the meantime, satellite operators plan to employ more bands of FR3 such as 13.85-14 GHz for target-satellite uplinks. Moreover, both the service link and feeder link shown in Fig.~\ref{Fig:Usecase} have employed FR3. Besides, advancing UAV communication in FR3 is a trend, because small drones prefer a smaller size antenna array, moreover, FR3 can confront the blockage effects that are regarded as the bottleneck of FR2. 

Following the existing radar deployment in FR3, ISAC becomes very promising as it allows the combining functionalities in radio systems. Moreover, employing multi-antenna and satellite systems for sensing exploits spatial diversity and ubiquitous connectivity, respectively, which enhances crowd, environment, and remote sensing shown in Fig.~\ref{Fig:Usecase}.

Considering the on-demand connectivity in terrestrial cellular networks for IMT, different FRs can be deployed in diverse scenarios categorized by distance. As shown in Fig.~\ref{Fig:Usecase}, we exemplify the usage scenarios of different FRs, in which FR3 is more suitable for medium-coverage deployments \cite{nokiabell}, such as urban, factory, and O2I scenarios. In comparison, FR1 can be deployed in rural and remote scenarios, while FR2 can be used for backhaul, vehicular, and hot-spot scenarios. It shows that FR3 could tailor a larger subset of use scenarios, compared to FR1 or FR2 each focusing more on specific use cases. 
More suggested scenarios are summarized in Table~\ref{tab1}, showing a higher use case flexibility in FR3.

\begin{figure}[!t]
  \centering
   {\includegraphics[width=3in]{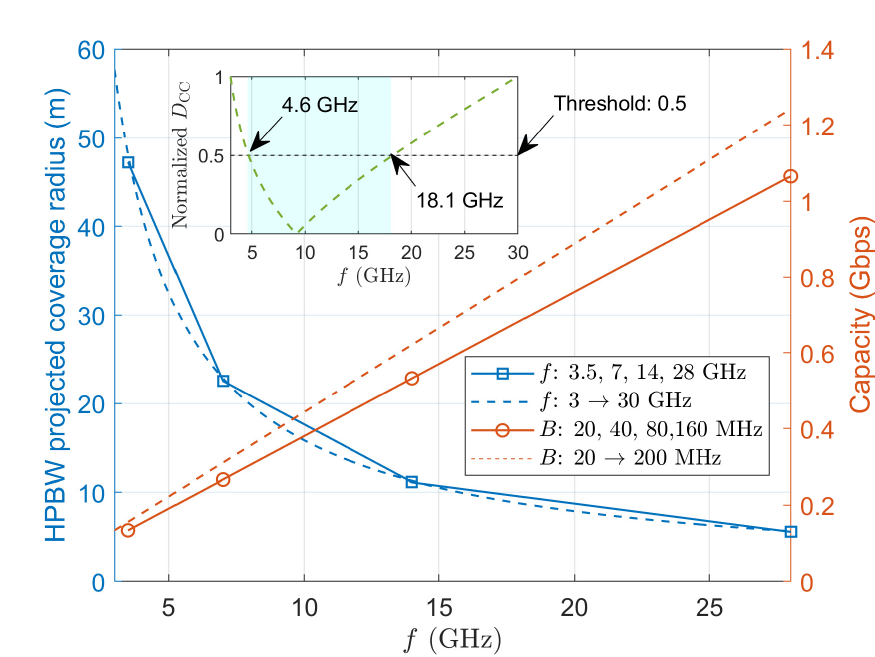}}
  \caption{A trade-off between beam coverage and capacity considering 3-30 GHz, where the radius is calculated by $H\tan(\rm{HPBW/2})$ ($H$ is the height difference) and capacity is determined by the bandwidth set to 20-200 MHz corresponding to frequency respectively.} 
  \label{cov-cap}
 \end{figure}

The high deployment flexibility is owing to its balance in coverage and capacity. An illustration is shown in Fig.~\ref{cov-cap}. Considering a UAV BS flying at 100~m, it is equipped with a uniform rectangular array (URA) like the ones in Table~\ref{tab1}. In this case, its beam coverage on the ground is determined by the HPBW, we then use the HPBW projected coverage radius to indicate the coverage level  (radius = $100\tan(\rm{HPBW/2})$). Channel capacity is determined by the bandwidth when we consider a free-space path loss and the same transmit power for different frequencies, resulting in the same signal-noise ratio (SNR). It is found that FR3 balances coverage and capacity better than other FRs. We use $N_{\rm CC}$, a normalized difference level between coverage and capacity, to indicate performance divergence. In the sub-illustration of Fig.~\ref{cov-cap}, the higher $N_{\rm CC}$ means a higher difference. By setting the threshold as 0.5, we found the band that can balance two metrics well located in 4.6--18.1 GHz. Thus, for systems operating at a frequency below 4.6 GHz or above 18.1 GHz, they should focus more on coverage or capacity, respectively. It shows that the majority of FR3 can provide decent performances for both. Moreover, HPBW suggests an angular resolution of sensing, therefore FRs 1--3 tend to be more favorable for presence detection, precise sensing such as human motions, and various sensing activities, respectively. 
We herein consider a clean channel, and the next section will elaborate on how similar the FR3 channel is to the other FRs in a multipath environment. 

\section{Radio Channel Characterization and Modeling in FR3 }
Exploring the potential of the wireless networks operating in FR3 first requires understanding the radio channel. Since the transmission schemes of FR3 systems become much more complicated, the overall radio channel observed by specified antenna arrays can be expected to be significantly different from the well-explored channels in 5G FR1 and FR2 bands. 


\subsection{Statistical Comparison of Beamforming across Bands}
To understand the key benefit of the FR3, we have to understand what is more valuable: similar channel characteristics or diverse channel characteristics? Similarity makes existing planning tools valid and enables information sharing across bands. Diverse channel characteristics make it possible to achieve diversity gains and enable maybe applications and technologies that were not there before. A key requirement to understand the value of FR3 is hence the comparison of FR3 with other frequencies.
The coexistence of multi-band networks not only enhances the coverage and capacity of conventional networks operating at a single frequency but also facilitates channel information sharing across multiple frequencies. For example, out-of-band spatial channel information collected normally via lower frequency signals can be utilized for coarse estimation of beam directions at higher frequencies. Such protocol helps to reduce beam training overhead for beam alignment and tacking. The success of out-of-band information aided beam search critically depends on the spatial channel similarity between lower- and higher-frequency channels. Many channel measurements at sub-6 GHz and 5G mmWave frequencies, i.e., 28 GHz and 39 GHz, do exist for visual inspection of power angular spectra (PASs). Recently, two metrics were proposed for fair comparison of multi-band PASs considering different beam patterns used at lower and higher frequencies \cite{pekka23}. The measured PAS of the propagation channel is filtered by the desired beam pattern resulting in the so-called \textit{beamformed PAS}. The first metric, i.e., \textit{PAS similarity percentage} (PSP), characterizes the total variation distance of normalized beamformed PASs. Here, the beamformed PAS is normalized to 
such that it can be interpreted as a probability distribution function (PDF). However, a low PSP does not mean that less useful spatial channel information can be transferred from lower to higher frequencies. The second metric is the ratio $R$ 
of the sum power of high-frequency beamformed PAS at desired beam directions extracted from lower- and higher-frequency radio channels.

\begin{figure}[!t]
\centering
\subfigure[][]{
    \label{Fig_psp}
    \includegraphics[width=3in]{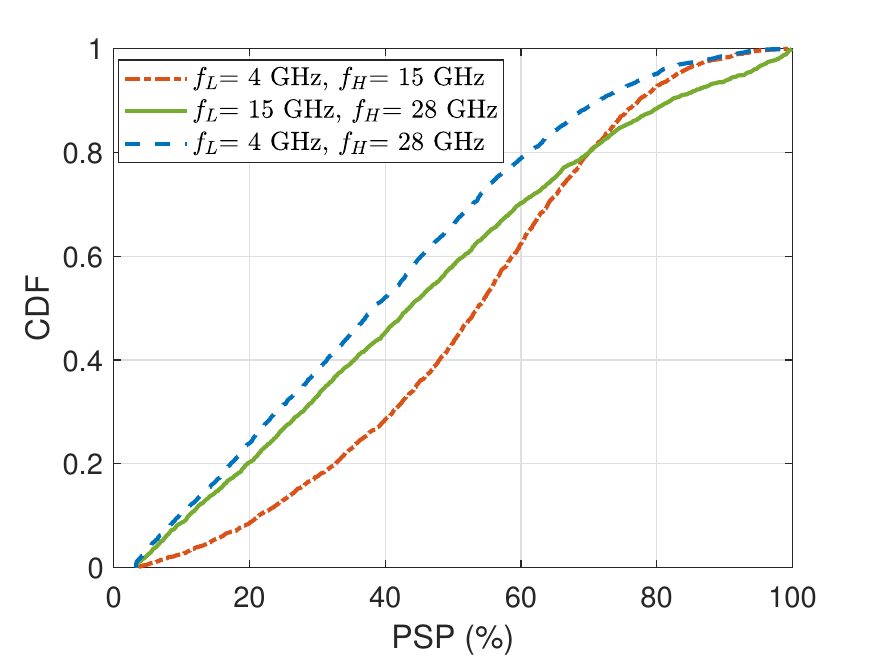}
}\\
\subfigure[][]{
    \label{Fig_rvalue}
    \includegraphics[width=3in]{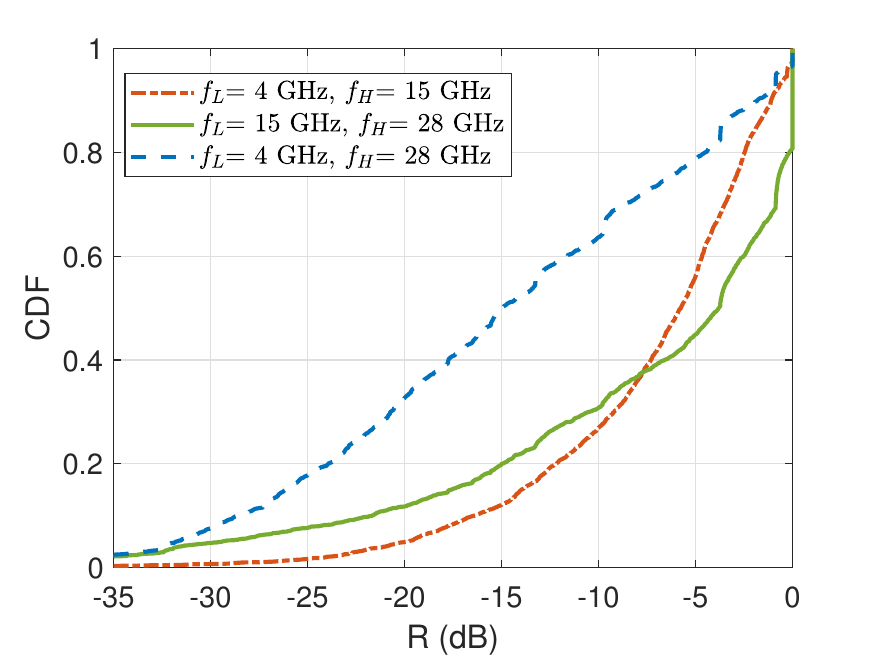}
}
\caption{CDFs of PSP and $R$ values for frequency band comparisons.}
\label{fig:Similarity}
\vspace{-6pt}
\end{figure}

The two metrics are now evaluated in a relevant FR3 deployment scenario, and FR3 is compared with both the lower FR1 and higher FR2 channels. Ray-tracing simulations in the airport check-in area were performed at 4, 15, and 28 GHz with over 2000 Tx-Rx links as detailed in \cite{pekka23} where 4, 15, and 28 GHz represent FR1, FR3, and FR2 respectively. Figs.~\ref{Fig_psp} and (b) compare the cumulative distribution functions (CDFs) of PSP and power ratio $R$ based on different frequency combinations, FR1-FR2, FR1-FR3, and FR3-FR2. Practical constraints on the antenna configurations are taken into consideration. Here, beam patterns of 4-, 16-, and 28-element half-wavelength-spaced uniform linear arrays (ULAs) are employed at 4, 15, and 28 GHz, respectively, to keep approximately the same antenna apertures at these frequencies. The improvement of angular resolution due to using larger antenna arrays leads to more potential beam directions extracted from higher-frequency radio channels. Consequently, the power ratio $R$, comparing ideal higher-frequency beamforming based on high-frequency beam scan, with higher-frequency beamforming with a lower-frequency beam scan, 
is less than or equal to 0 dB. 
The similarity levels between 4 GHz and 15 GHz radio channels, as well as 15 GHz and 28 GHz channels, are much higher than the case of 4 GHz and 28 GHz channels. In particular, such an increase in $R$ value becomes more significant regarding beam pointing direction-based metric. By comparing spatial channel characteristics, a network deployed in the FR3 can explore more accurate spatial channel information from sub-6 GHz channels and provide more useful direction information for mmWave communications.

\subsection{Key Aspects of FR3 Channel Models}
Despite widely used channel models by the industry throughout the 4G and 5G communications can support up to 100 GHz channel simulations \cite{3gpp901}, the model parameters were mainly extracted from 
field measurements in FR1 and FR2 bands. Because of a lack of measurement results in the 7--24 GHz range, new measurements are needed to provide a statistically significant number of measurement points that could form a reliable basis for geometry-based stochastic channel models (GSCMs). The GSCM includes the concept of cluster and each cluster is a set of plane wave components or multipath components (MPCs) having slightly different propagation parameters, including path gain, delay, directions of arrival and departure, Doppler shift, and cross-polarization ratio. The statistics of these parameters will be first updated based on new measurements accordingly. Moreover, a communication system operates in the radio channel observed by the Tx and Rx antennas, rather than the propagation channel. Thus, new channel characteristics need to be included in the light of emerging transmission schemes. As aforementioned, upper mid-band systems are promising to be equipped with larger antenna arrays, which not only provide more DoFs, 
but also introduce the near-field effect. Additional model components and parametrization are needed to accurately capture these effects under the existing GSCM framework.

A generalized channel model for upper mid-band systems needs to be constructed that combines both terrestrial and non-terrestrial channels. For example, the channel models in \cite{3gpp901} and \cite{3gpp38811} were developed purely for terrestrial and non-terrestrial networks, respectively. The main difference is caused by the elevated, e.g., satellite, high-altitude platform as IMT BS (HIBS), and UAV. Our air-to-ground (A2G) channel measurement results show that the ITU diffraction model has good agreements at 1 GHz and 4~GHz, while an additional height-dependent parameter is essential to describe the extra loss due to diffraction and scattering at 12 GHz and 24 GHz \cite{cesar20}. For satellite/HIBS channels, atmospheric effects including atmospheric absorption, rain and cloud attenuation, scintillation, and Faraday rotation cannot be ignored as well.

Another challenging problem of the existing stochastic models is that they cannot be directly used for ISAC channel and environment generation. It becomes important to use an integrated model that fully considers channel geometry with correct mapping of target scattering properties in the 3D global coordinate system. Thus, a hybrid modeling methodology is expected to be used for ISAC channel modeling. In particular, deterministic characterization of the scattering properties of typical objects (e.g., people, vegetation, and buildings) in the FR3 band requires extensive site-specific measurements and electromagnetic simulations.

\section{Key Techniques in FR3}
In this section, we highlight the three key technologies for 6G 
and specify the dedicated design requirements when considering FR3 deployment. A case study to verify the benefit of using FR3 in the RIS is conducted. 

\subsection{NTN}
Key nodes in NTN including UAVs, HIBS, and LEO satellites have a common size-limited feature, unlike terrestrial stations. We first show the potential of FR3 in NTN and then discuss the future focuses towards efficient implementations. 

\subsubsection{Potentials} 
FR3 encompasses many band windows dedicated to space-earth, UAS, and deep-space communications \cite{3gppfr3}. Consequently, the primary components of NTN can effectively harness these frequency bands. For UAVs that fly in low altitudes ($\le$1~km), an urgent need is to constantly connect with ground cellular BSs so that reliable links can be established. However, using either analog or digital beamforming, the low side-lobe gains or 
expensive up-tilted beams in FR1 and FR2 respectively hinder this connectivity. FR3-enabled 3D coverage realized potentially by hybrid beamforming could play an important role in supporting UAVs, in an integrated terrestrial network (TN) and NTN framework. 

HIBS and LEO satellites are generally used in 20-50~km and 300-2000~km, respectively, resulting in almost pure line-of-sight (LoS) channels. However, the high altitudes suggest that the operating frequency cannot be too high, so mmWave bands become gloomy. Besides, due to the size and payload limitations, equipping nodes with a large number of antenna elements in sub-6~GHz becomes impossible. Therefore, FR3-based arrays with smaller sizes can serve as promising implementation solutions for these space/aerial nodes. 



\subsubsection{Focuses}
Current commercial satellite services, exemplified by Starlink, typically utilize frequencies ranging from 10.7 to 12.7 GHz for the downlink and 14.0 to 14.5 GHz for the uplink. Additionally, FCC has recently designated a band of 18.1 to 18.6 GHz for satellite exploration, highlighting its significance for satellite communication. As shown in Fig.~\ref{cov-cap}, considering the size-limited feature of NTN nodes, FR3 provides a well-balanced coverage and capacity in a UAV scenario. However, atmospheric and rain attenuation that can be accumulated along long distances should be further characterized in FR3, because a large range in FR3 may exhibit frequency-dependent characteristics. Furthermore, it is important to model complex multipath structures and ensure their consistency with FR1 and FR2. Lastly, since a clean environment and multi-beam transmission lead to inter-cell and intra-cell interference, respectively, interference modeling and management is another key issue in NTNs.

\subsection{RIS}

\subsubsection{Potentials} 
RIS has attracted much interest thanks to its ability to control the radio signals between transmitter and receiver. For instance, a liquid-crystal embedded reflectarray antenna for
6G FR3 is designed in \cite{antennafr3}. 
The number of elements plays an important role in the RIS-aided communication performance. To this end, we conduct a comparison among different FRs considering the same size of the array, where the RIS is put in the middle of the transmitter and receiver. The detailed parameters are as follows. Considering an indoor environment, the 3D coordinates of the transmitter, RIS, and receiver are $(0,0,3), (2.5, 5, 3), (5,0,1.5)$~m, respectively. 
 For the same RIS array size, FR3 results in a higher number of half-wavelength-spaced elements compared to the lower FR1. For a square array, this means that the array of a $10\times10$ RIS at 15 GHz, resulting in $20\times20$~cm area. In the same area, the RIS becomes $2\times2$ and $18\times18$ at 3.5~GHz and 28 GHz, respectively. Besides, the transmit power ranges from -10 to 40~dBm, and the power consumption of phase shifters in RIS is set as 7.8~mW, 4.5~mW, and 1.5~mW for 6-bit, 4-bit, and 3-bit \cite{rispower} at 3.5~GHz, 15 GHz and 28 GHz, given the higher-resolution requirement when having fewer elements. 

\begin{figure}[!t]
  \centering
   {\includegraphics[width=3in]{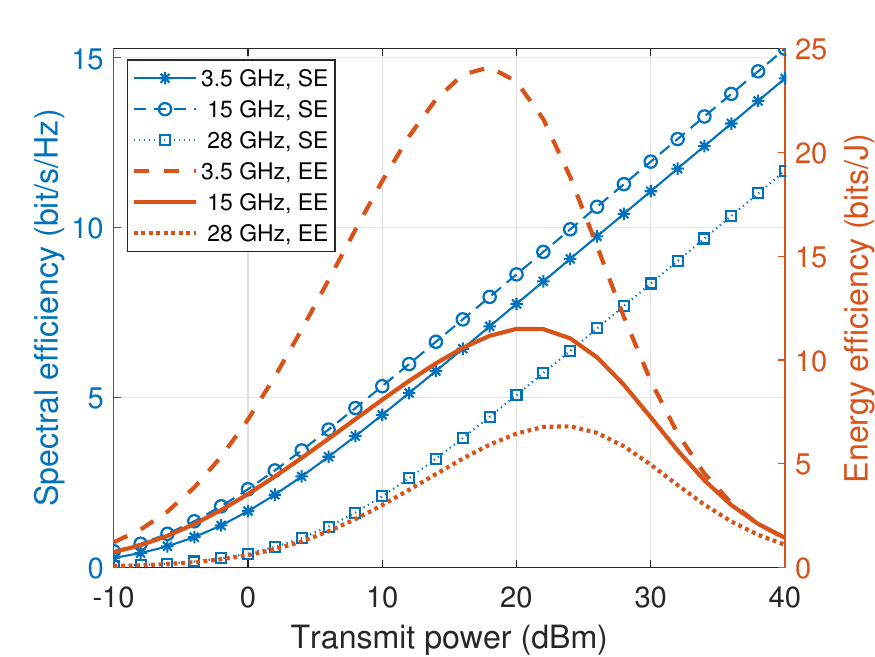}}
  \caption{SE and EE at 3.5, 15, and 28 GHz as a representative in each FR, where we consider the fixed size and location of RIS in an indoor environment.}
  \label{riscom}
 \end{figure}

In this case study, the spectral and energy efficiency (SE and EE) are analyzed, where SE is calculated based on the SNR, and EE calculation considers the total power consumption consisting of transmit power, user power, and the overall consumption of RIS elements. As shown in Fig.~\ref{riscom}, it is observed that the SE at 15~GHz is higher than those of the other two frequencies. On average, the SE is improved by 0.87 and 3.53~bit/s/Hz compared to 3.5~GHz and 28~GHz, respectively. The EE performances considering 20 users with the power of 10~dBm become worse than 3.5~GHz but better than 28~GHz, given the fact that 4, 100, and 324 elements are used in the RIS at 3.5~GHz, 15~GHz, and 28~GHz, respectively. However, the EE performance is always better than mmWave bands, and close to that at 3.5 GHz in the high SNR regime. Overall, when jointly considering the SE and EE performances, FR3 is still promising. 

\subsubsection{Focuses} We can find that the power consumption of phase shifters directly impacts the performance. Therefore, one focus will be the hardware testing and modeling in FR3, such as power amplifiers, phase shifters, and array configurations. Then, when using multi-antenna systems in other techniques, for instance, RIS for sensing, near-field effect led by large arrays may benefit sensing by providing more angular information, however, the coupling effect in communication should be well modeled.  
Besides, FR3 has a wide bandwidth and relatively rich multipath, combined with large arrays. These features open the opportunity for the hybrid MIMO, through balancing the benefits of LoS phased array beamforming with many antennas, and digital processing of multiple streams for spatial multiplexing and exploiting multipath.

 \begin{figure*}[!t]
  \centering
   {\includegraphics[width=5in]{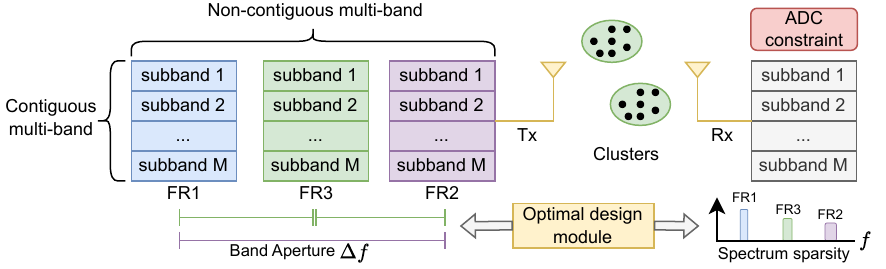}}
  \caption{An illustration of multi-band ISAC system considering aggregation of FRs.}
  \label{multiband}
 \end{figure*}
\subsection{ISAC}
\subsubsection{Potentials} 
Since there are complete hardware components for sensing purposes in the communication system, it is more cost-efficient to incorporate sensing functions into the communication procedure. However, sensing performance highly relies on spectrum resources, which are constrained by the bandwidth of communication signals \cite{multi-band}. To overcome spectrum and hardware limitations, a promising direction is the use of 
multi-band ISAC to improve the sensing resolution. As shown in Fig.~\ref{multiband}, we illustrate the multi-band ISAC system. Based on communication systems with features of limited bandwidth and analog-digital converter (ADC) constraints, aggregating 
intra-FR and inter-FR bands results in contiguous and non-contiguous multi-band schemes, respectively. 

Considering the coexistence of 5G NR and future 6G NR, the non-contiguous strategy provides a wider band aperture $\Delta f$, leading to a higher delay resolution. Moreover, FR-level aggregation ensures spectrum sparsity so that advanced signal processing such as compressed sensing can be employed, reducing the requirement of ADC. However, the frequency dependence of the channel becomes significant when using too large $\Delta f$ because of the changes in the multipath structure and frequency-dependency of Doppler shifts. Thus, as a bringing band, FR3 plays an important role in aggregating FR1 or FR2, rather than aggregation between FR1 and FR2. Our results in Fig.~\ref{fig:Similarity} show both PSP and $R$ have lower values for the case of $f_L=4$~GHz and $f_H=28$~GHz, which indicates the frequency dependence is higher than FR3-based comparisons.

\subsubsection{Focuses} 
Multi-band systems focus on the practical coexistence of FRs, however, the assumption of frequency independence needs to be well verified. Besides, considering the three dimensions including space, time, and frequency to implement ISAC, the focus also lies in comprehensive channel characteristics in FR3, where multipath structure in the delay, Doppler, and power domains and corresponding spatial non-stationarity when using ELAAs should be clarified. 
Finally, embedding FR3 and ISAC channel measurements and modeling into the existing 3GPP channel modeling framework \cite{3gpp901} has become a critical study item in Release 19.  

A system-level implementation of ISAC in FR3 requires optimal design trade-offs. Balancing arrays, bandwidths, and frames, needs advanced optimization algorithms to improve the range-angle-Doppler resolution. Moreover, hardware imperfections need to be compensated by optimal deployment schemes such as multi-static ones. Lastly, the EE of ISAC systems is another focus, because there is a contradiction between the radar and communication subsystems, where the former uses more power to increase detection range, while the latter prefers to transmit more bits using less power.





\section{Challenges}
In this section, we point out research challenges from channels and spectrum to hardware and realistic implementations. 
\subsection{Wireless Channel Modeling}
As the FR3 can be used in both NTN and TN, an integrated channel model is required, taking diverse propagation environments ranging from rich-scattered urban to scattering-free space, distinct mobility, and frequency dependence into account. Then, to empower RIS-aided communication and sensing, radio channel models should fully incorporate both physical scattering from the environment and artificial reflections controlled by RIS. Lastly, a unified channel model should be developed for ISAC, since the sensing channel aims to locate scatterers in a deterministic way while the communication channel is described in probability distributions such as Rayleigh and Rician, which is a stochastic way. The latter is extremely challenging in a multi-band context.


\subsection{Spectrum Coexistence and Aggregation}
Future 6G NR will use more allocated FRs, aiming to provide diverse connectivity options to upgrade dual connectivity in 5G. However, a well-designed spectrum management scheme is required, which needs a thorough evaluation of coverage, throughput, efficiency, and implementation cost of the corresponding spectrum users. A multi-dimensional spectrum
management approach is needed, that monitors if end users
are indeed making the most value out of their assigned spectrum. Then, spectrum aggregation by including FR3 enables high-resolution sensing, however, it also poses a challenge in the trade-off design of balancing hardware limitation, sensing performance, and communication efficiency. 

\subsection{Adaptive Architecture Design}
 To smoothly operate radio frequency front-ends to 7--24~GHz, new transceivers need to be designed, in terms of up/down converters, ADC considering wider bandwidth, and power amplifiers that can have linear output in FR3. Moreover, the transceiver architecture designed in FR1 and FR2 should be adapted to FR3. Moreover, FR3 will be a golden band for hybrid beamforming architecture, as it is helpful to balance spatial multiplexing in FR1 and reduce beamformer complexity and power consumption in FR2. The pivotal challenge is to balance optimization-based and learning-based designs for the hybrid transceiver architecture.  

\subsection{Realistic Applications Implementation}
Existing studies and usage show that FR3 can be used in cellular systems (potentially in 7--15 GHz \cite{nokiabell}) and inter-satellite systems that can be operated in 22.55--23.55 GHz \cite{wrc23final} to consider fair and targeted coverage, respectively. This versatility allows FR3 to facilitate both broad and directional communications when using the lower part and higher part of 7-24 GHz. Considering the applications in terms of their functionality and environment, an application-specific instruction set processor (ASIP)-like implementation is promising but also challenging, as the FR3 band is more flexible and programmable than the other bands. Designing such an implementation entails further categorization of FR3 applications and their specific requirements. 

\balance

\section{Conclusion}
In the article, we provided an overview of the radio spectrum used in the context of 6G networks, highlighting the immense potential of FR3 spanning the range of 7--24 GHz. From antenna-friendly dimensions, and favorable propagation, to increased sensing accuracy, we outlined the fundamental advantages of FR3. By comparing channel characteristics at 4, 15, and 28 GHz, ray-tracing simulations show that FR3 can obtain more accurate spatial channel information from sub-6 GHz channels and provide more useful direction information for mmWave bands. We then investigated the benefits of using FR3 in three emerging techniques regarding NTN, RIS, and ISAC. Additionally, by evaluating the RIS-assisted communication, we revealed SE enhancements at 15 GHz when considering the same RIS size, showing improvements of 0.87 and 3.53 bit/s/Hz compared to 3.5 GHz and 28 GHz, respectively. Finally, we outlined challenges for the future deployment of FR3, in terms of channel modeling, spectrum management, hardware compatibility, and on-demand implementation.

\section*{Acknowledgment}
This work is supported by the iSEE-6G SNS-JU project under the European Union’s Horizon Europe research and innovation programme with Grant Agreement no. 101139291.


\ifCLASSOPTIONcaptionsoff
  \newpage
\fi

\bibliographystyle{IEEEtran}
\bibliography{IEEEabrv,Reference}
\vspace{0.5cm}
\begin{IEEEbiographynophoto}{Zhuangzhuang Cui}
 received the Ph.D. degree in Electrical Engineering from the Beijing Jiaotong University, China, in 2022. Currently, he is a postdoctoral researcher with the Department of Electrical Engineering, KU Leuven, Belgium. His research interests include channel modeling, UAV, ISAC, and NTN. 
\end{IEEEbiographynophoto}
\vspace{-0.5cm}
\begin{IEEEbiographynophoto}{Peize Zhang} received the Ph.D. degree (Hons.) in Electrical Engineering from Southeast University, China, in 2022. He is a postdoctoral researcher in the 6G Flagship programme with the Centre for Wireless Communications (CWC), University of Oulu, Finland. His current research interests focus on mmWave/THz communication systems.
\end{IEEEbiographynophoto}
\vspace{-0.5cm}
\begin{IEEEbiographynophoto}{Sofie Pollin} received the
Ph.D. degree (Hons.) from KU Leuven in 2006. She is currently a Full Professor with the Department of Electrical Engineering at KU Leuven, Belgium, leading the Networked Systems group affiliated with WaveCoRE. She is also a Principle Scientist at imec, Belgium. 
Her research interests are cell-free networks, ISAC, and NTN. 
\end{IEEEbiographynophoto}

\end{document}